\documentclass[aps,prl,twocolumn,preprintnumbers,superscriptaddress,groupedaddress,nofootinbib]{revtex4}  
\usepackage{bm}
\usepackage{amssymb}
\usepackage{amsmath,shuffle}

\allowdisplaybreaks

\usepackage{tikz}
 \usetikzlibrary{calc} \usetikzlibrary{patterns} \usetikzlibrary{decorations.pathreplacing}
\usetikzlibrary{decorations.markings} \usetikzlibrary{decorations.pathmorphing} \usetikzlibrary{positioning}

\def\eqref#1{(\ref{#1})}
\def\beq{\begin{equation}}
\def\eeq{\end{equation}}
\def\tr{\text{Tr}}
\def\PsiF{{\mathbb V}}

\def\a{\alpha}
\def\b{{\beta}}
\def\g{{\gamma}}
\def\d{{\delta}}

\def\l{\lambda}

\def\t{{\theta}}

\def\p{{\partial}}
\def\({\left(}
\def\){\right)}
\def\cF{{\cal F}}
\def\cW{{\cal W}}

\def\cA{{\cal A}}

\def\cK{{\cal K}}
\def\cN{{\cal N}}
\def\bA{{\mathbb A}}
\def\bW{{\mathbb W}}
\def\bF{{\mathbb F}}
\def\Wg{\bW}
\def\Fg{\bF}

\begin{document}

\preprint{AEI--2015--005, DAMTP--2015--5}

\title{A solution to the non-linear equations of ${\bm D} {\bm =} {\bm 1}{\bm 0}$ super Yang--Mills theory}

\author{Carlos R. Mafra}
\email{crm66@damtp.cam.ac.uk}
\affiliation{DAMTP, University of Cambridge, Wilberforce Road, Cambridge, CB3 0WA, UK}
\author{Oliver Schlotterer}
\email{olivers@aei.mpg.de}
\affiliation{Max--Planck--Institut f\"ur Gravitationsphysik, Albert--Einstein--Institut, 14476 Potsdam, Germany}

\date{\today}

\begin{abstract}
In this letter, we present a formal solution to the non-linear field equations of ten-dimensional super
Yang--Mills theory. It is assembled from products of linearized superfields which have been
introduced as multiparticle superfields in the context of superstring perturbation theory. Their
explicit form follows recursively from the conformal field theory description of the gluon multiplet in
the pure spinor superstring. Furthermore, superfields of higher mass dimensions are defined and their
equations of motion spelled out.
\end{abstract}

\maketitle

\section{\label{sec:intro}Introduction}

Super Yang--Mills (SYM) theory in ten dimensions can be regarded
as one of the simplest SYM theories, its spectrum contains just the gluon and gluino, related by sixteen supercharges.
However, it is well-known that its dimensional reduction gives rise to various 
maximally supersymmetric Yang--Mills theories in lower dimensions,
including the celebrated $\cN = 4$ theory in
$D=4$ \cite{SYM}.
Therefore a better understanding of this theory propagates a variety
of applications to any dimension $D\leq 10$.

In a recent line of research \cite{npttree,t1loop}, scattering amplitudes of ten-dimensional SYM have been determined
and simplified using so-called multiparticle superfields \cite{EOMBBs}. They represent entire tree-level subdiagrams
and build up in the conformal field theory (CFT) on the worldsheet of the pure spinor superstring \cite{psf} via operator product expansions
(OPEs). Multiparticle superfields satisfy the linearized field equations with the addition of contact terms, i.e.~inverse off-shell
propagators. In this letter we demonstrate that these off-shell modifications can be resummed to
capture the non-linearities in the SYM equations of motion. The generating series of multiparticle
superfields as seen in (\ref{Lie}) is shown to solve the non-linear field equations spelled out in (\ref{EOMs}).

We also define superfields of arbitrary mass dimension and reduce their non-linear expressions to the
linearized superfields of lower mass dimensions. This framework simplifies
the expressions of kinematic factors in higher-loop scattering amplitudes, including the
$D^6R^4$ operator in the superstring three-loop amplitude \cite{3loop}.

\section{\label{sec:review} Review of ten-dimensional SYM}

The equations of motion of ten-dimensional SYM theory can be described
covariantly in superspace by defining
supercovariant derivatives \cite{siegel, wittentwistor}
\begin{equation}
\label{covder}
\nabla_\a \equiv D_\a - \bA_\a(x,\t) ,\quad \nabla_m \equiv \p_m - \bA_m(x,\t) .
\eeq
The connections $\bA_\a$ and $\bA_m$ take values in the Lie algebra
associated with the non-abelian Yang--Mills gauge group. The derivatives are taken with respect
to ten-dimensional superspace coordinates $(x^m,\t^\a)$ with vector and spinor
indices $m,n=0, \ldots,9$ and $\a,\b=1, \ldots, 16$ of the Lorentz group. The
fermionic covariant derivatives
\begin{equation}
\label{deralg}
D_\a \equiv \p_\a + \tfrac{1}{2}(\g^m\t)_\a\p_m , \quad \{D_\a,D_\b\} = \g^m_{\a\b}\p_m
\eeq
involve the $16\times 16$ Pauli
matrices $\g^m_{\a\b}=\g^m_{\b\a}$ subject to
the Clifford algebra
$\g^{(m}_{\a\b}\g^{n)\b\g} = 2\eta^{mn}\d_\a^\g$, and the convention for (anti)symmetrizing indices does not include~$\tfrac{1}{2}$.

The connections in (\ref{covder}) give rise to field-strengths
\beq
\label{FS}
\bF_{\a\b} \equiv \{\nabla_\a,\nabla_\b\} - \g^m_{\a\b}\nabla_m, \quad
\bF_{mn} \equiv - [\nabla_m,\nabla_n]\,  .
\eeq
One can show that the constraint
equation $\bF_{\a\b}=0$ puts the superfields on-shell, and Bianchi identities
lead to the non-linear equations of motion \cite{wittentwistor},
\begin{align}
 \big\{ \nabla_{\alpha} , \nabla_{\beta} \big\} &= \gamma^m_{\alpha \beta} \nabla_m \cr
\big[ \nabla_\alpha , \nabla_m\big] &= - (\gamma_m \bW)_\alpha  \cr
\big\{ \nabla_{\alpha} ,\bW^\beta \big\} &= \tfrac{1}{4} (\gamma^{mn})_{\alpha}{}^{\beta} \bF_{mn} \cr
\big[\nabla_{\alpha}, \bF^{mn} \big] &= \big[ \nabla^{[m} ,(\gamma^{n]} \bW)_\alpha \big] \ .
\label{EOMs}
\end{align}
In the subsequent, we will construct an explicit solution for the superfields $\bA_\a$, $\bA_m$,
$\bW^\a$ and $\bF^{mn}$ in (\ref{EOMs}). The main result is furnished by the generating series in (\ref{Lie}) whose constituents will be introduced in the next section.

\section{\label{sec:EOMBBs} Linearized multiparticle superfields}

In perturbation theory, it is conventional to study solutions
$A_\a, A_m,\ldots$ of the \textit{linearized} equations of motion
\begin{align}
\big\{ D_{(\a} , A_{\beta)} \big\} &= \gamma^m_{\a \beta} A_m \cr
\big[ D_\a , A_m\big] &= k_mA_{\a} + (\gamma_m W)_\a \cr
\big\{ D_{\a} , W^\beta \big\} &= \tfrac{1}{4} (\gamma^{mn})_{\a}{}^{\beta} F_{mn} \cr
\big[D_{\a}, F^{mn} \big] &= k^{[m} (\gamma^{n]} W)_\a \ .
\label{linEOMs}
\end{align}
Their dependence on the bosonic coordinates $x$ is described by plane waves $e^{k\cdot x}$ with
on-shell momentum $k^2=0$. In a gauge where $\theta^\alpha A_\alpha=0$, the $\theta$ dependence is known in terms of fermionic power series
expansions from \cite{harnad,tsimpis} whose coefficients contain gluon polarizations and gluino wave functions.

As an efficient tool to determine and compactly represent scattering amplitudes in SYM and string
theory, multiparticle versions of the linearized superfields have been constructed in \cite{EOMBBs}. They
satisfy systematic modifications of the linearized equations of motion \eqref{linEOMs}, and their
significance for BRST invariance was pointed out in \cite{jonas}.
For example,
their two-particle version
\begin{align}
A^{12}_\a &\equiv - \tfrac{1}{2}\bigl[ A^1_\a (k^1\cdot A^2) + A^1_m (\g^m W^2)_\a - (1\leftrightarrow 2)\bigr]\cr
A^{12}_m &\equiv  \tfrac{1}{2}\bigl[ A_1^p F^2_{pm} - A^1_m(k^1\cdot A^2) + (W^1\g_m W^2) - (1\leftrightarrow
2)\bigr]\cr
W_{12}^\a &\equiv \tfrac{1}{4}(\g^{mn}W_2)^\a F^1_{mn} + W_2^\a (k^2\cdot A^1) - (1\leftrightarrow 2) \label{twopart} \\
F^{12}_{mn} &\equiv F^2_{mn}(k^2 \cdot A^1)+ \tfrac{1}{2} F^{2}_{[m}{}^p F^{1}_{n]p}  \notag \\
& \  \ + k^1_{[m}(W^1 \gamma_{n]} W^2) - (1\leftrightarrow 2) \,, \notag
\end{align}
can be checked via \eqref{linEOMs} to satisfy the following two-particle equations of motion:
\begin{align}
\label{twoEOM}
\{ D_{(\a}, A^{12}_{\b)} \}&= \g^m_{\a\b}A^{12}_m + (k^1\cdot k^2)(A^1_\a A^2_\b + A^1_\b A^2_\a) \\
[D_\a, A_{12}^m] &= \g^m_{\alpha \beta} W_{12}^{\beta} \!+\! k_{12}^m A^{12}_\a \!+\! (k^1\cdot k^2)(A^1_\a A_2^m \!-\! A^2_\a A_1^m)\cr
\{D_\a, W^\b_{12} \}&= \tfrac{1}{4}(\g^{mn})_\a{}^\b F^{12}_{mn} + (k^1\cdot k^2)(A^1_\a W_2^\b - A^2_\a W^\b_1)\cr
[D_\a, F^{12}_{mn} ]&= k^{12}_m (\g_n W^{12})_\a - k^{12}_n (\g_m W^{12})_\a\cr
& \! \! \!\! \! \! \! \! \! + (k^1\cdot k^2)\big(A^1_\a F^2_{mn}  +  A^{1}_{[n} (\g_{m]} W^2)_\a - (1\leftrightarrow 2)\big)\, .\notag
\end{align}
The modifications as compared to the single-particle case \eqref{linEOMs} involve the
overall momentum $k_{12} \equiv k_1+k_2$ whose propagator is generically off-shell, $k_{12}^2= 2(k_1\cdot k_2) \neq
0$.

The construction of the two-particle superfields in (\ref{twopart}) is guided by string theory methods. In the pure spinor formalism \cite{psf}, the insertion of a gluon multiplet state on the boundary of an open string worldsheet is described by the integrated vertex operator
\beq
\label{Uvertex}
U^i \equiv \p\t^\a A^i_\a + \Pi^m A^i_m + d_\a W_i^\a + \tfrac{1}{2} N^{mn}F^i_{mn} \ .
\eeq
Worldsheet fields $[\p\t^\a,\Pi^m,d_\a, N^{mn}]$ with conformal weight one and well-known OPEs
are combined with linearized superfields associated with particle label $i$.
The multiplicity-two superfields in (\ref{twopart}) are obtained from the coefficients of the
conformal fields in the OPE \cite{EOMBBs}
\begin{align}
U^{12} & \equiv  - \oint (z_{1}-z_{2})^{\alpha' k^1 \cdot k^2} U^1(z_1) U^2(z_2)
\label{oint}\\
&=
\p\t^\a A^{12}_\a + \Pi^m A^{12}_m + d_\a W^\a_{12} + \tfrac{1}{2} N^{mn}F^{12}_{mn}\,,\notag
\end{align}
where $\alpha'$ denotes the inverse string tension, and total derivatives in the insertion points $z_{1},z_{2}$ on the worldsheet have been discarded in the second line. The contour integral in (\ref{oint}) isolates the singular behaviour of the $U^i$ w.r.t.~$(z_1-z_2)$ which translates into propagators $k_{12}^{-2}$ of the gauge theory amplitude after performing the $\alpha' \rightarrow 0$ limit. In other words, OPEs in string theory govern the pole structure of tree-level subdiagrams in SYM theory obtained from the point-particle limit.

The CFT-inspired two-particle prescription (\ref{twopart})
can be promoted to a recursion leading to superfields of arbitrary multiplicity, see (3.54), (3.56) and (3.59) of \cite{EOMBBs}. Their equations of motion are observed to generalize along the lines of
\begin{align}
\label{DAthree}
\bigl\{D_{(\a}, A^{123}_{\b)}\bigr\} &= \g^m_{\a\b}A^{123}_m+ (k^{12}\cdot k^3)\bigl[A^{12}_\a A^3_\b - (12\leftrightarrow 3)\bigr] \notag\\
&\!\! \! \! + (k^1\cdot k^2)\bigl[A^1_\a A^{23}_\b + A^{13}_\a A^2_\b - (1\leftrightarrow 2)\bigr]
\end{align}
for suitable definitions of $A_\a^{123}$ and $A_m^{123}$, see (3.17), (3.19) and (3.29) of \cite{EOMBBs}.

Multiparticle superfields can be arranged to satisfy kinematic analogues of the Lie algebraic Jacobi relations among structure constants, e.g. $A_\alpha^{123}+A_\alpha^{231}+A_\alpha^{312}=0$. They therefore manifest the BCJ duality \cite{BCJ1} between color and kinematic degrees of freedom in scattering amplitudes, see \cite{BCJ2} for a realization at tree-level. Together with the momenta $k_{12\ldots j}\equiv k_1+k_2+\ldots +k_j$ in their equations of motion, this suggests to associate them with
tree-level subdiagrams shown in the subsequent figure \cite{EOMBBs}:
\begin{center}
\tikzpicture[scale = 0.6, line width=0.30mm, xshift=-1cm]
 \draw (0,0) -- (-1,1) node[left]{$2$};
 \draw (0,0) -- (-1,-1) node[left]{$1$};
 \draw (0,0) -- (5,0);
 \draw (0.5,-0.4) node{$k_{12}$};
 \draw (1,0) -- (1,1) node[left]{$3$};
 \draw (1.6,0.4) node{$k_{123}$};
 \draw (2.2,0) -- (2.2,1) node[left]{$4$};;
 \draw (3, 0.5) node{$\ldots$};
 \draw (4,0) -- (4,1) node[left]{$p$};;
 \draw (4.7,-0.4) node{$k_{12...p}$};
 \draw (5.5,0) node{$\ldots$};
 \draw (8.4,0) node{$\leftrightarrow \ \ \left\{ \begin{array}{r} A_{\alpha}^{123\ldots p}
 \\ A_{m}^{123\ldots p} \\ W^{\alpha}_{123\ldots p} \\ F_{mn}^{123\ldots p}
 \end{array} \right.$};
\endtikzpicture
\end{center}

\vspace{-0.3cm}

\noindent
The cubic-graph organization of superfields already accounts for the quartic vertex in the bosonic Feynman rules of SYM. This ties in with the string theory origin of $n$-particle tree-level amplitudes where each contribution stems from $n-3$ OPEs.


\noindent
{\bf Berends--Giele currents:} 
As a convenient basis of multiparticle fields $K_B \in \{ A^{B}_\alpha, A^m_{B} ,
W_{B} ^\alpha, F_{B} ^{mn} \}$ with multiparticle label $B=12\ldots p$,
we define Berends--Giele currents
${\cal K}_B \!\in\! \{  \cA^{B}_\alpha, \cA^m_{B} , \cW_{B} ^\alpha, \cF_{B} ^{mn} \}$, e.g. ${\cal
K}_{1} \equiv  K_{1}$ and \cite{EOMBBs}
\begin{align}
\label{BGexpl}
{\cal K}_{12} \equiv \frac{K_{12}}{s_{12}} \ ,
\quad {\cal K}_{123} \equiv \frac{K_{123}}{s_{12}s_{123}} +\frac{K_{321}}{s_{23}s_{123}}
\end{align}
with generalized Mandelstam invariants $s_{12\ldots p} \equiv \tfrac{1}{2} k_{12\ldots p}^2$.
Berends--Giele currents ${\cal K}_B$ are defined to encompass all tree subdiagrams
compatible with the ordering of the external legs in $B$. The propagators $s^{-1}_{i\ldots j}$ absorb the appearance of explicit momenta in the contact terms of the equations of motion such as (\ref{twoEOM}) and (\ref{DAthree}).

As illustrated in the following figure, the
three-particle current in (\ref{BGexpl}) is assembled from the $s$- and $t$-channels of a color-ordered four-point
amplitude with an off-shell leg (represented by $\ldots$):

\begin{tikzpicture} [scale=0.6, line width=0.30mm]
\draw (-2.9,0) node{${\cal K}_{123} \  = $};
\draw (0,0) -- (-1,1) node[left]{$2$};
\draw (0,0) -- (-1,-1) node[left]{$1$};
\draw (0,0) -- (1.8,0);
\draw (0.5,0.3) node{$s_{12}$};
\draw (1,0) -- (1,1) node[right]{$3$};
\draw (1.65,-0.3) node{$s_{123}$};
\draw (2.75, 0) node{${\cdots} \  \ + $};
\scope[xshift=-0.4cm]
\draw (5.5,0) -- (4.5,1) node[left]{$3$};
\draw (5.5,0) -- (4.5,-1) node[left]{$2$};
\draw (5.5,0) -- (7.3,0);
\draw (6,-0.3) node{$s_{23}$};
\draw (6.5,0) -- (6.5,-1) node[right]{$1$};
\draw (7,0.3) node{$s_{123}$};
\draw (7.85, 0) node{${\ldots} $};
\endscope
\end{tikzpicture}

\noindent In contrast to the bosonic Berends--Giele currents in \cite{Berends:1987me}, the currents ${\cal K}_B$ manifest maximal supersymmetry, and their construction does not include any quartic vertices. A closed
formula at arbitrary multiplicity \cite{Broedel:2013tta, EOMBBs} involves the inverse of the momentum
kernel\footnote{The momentum kernel is defined by \cite{momKern}
\[
S[\sigma(2,3,\ldots,p) | \rho(2,3,\ldots,p)]_1 = \prod_{j=2}^{p} (s_{1,j_\sigma} + \sum_{k=2}^{j-1} \theta(j_\sigma,k_\sigma) s_{j_\sigma,k_\sigma})
\]
and depends on reference leg $1$ and two permutations $\sigma,\rho \in S_{p-1}$ of additional $p-1$ legs $2,3,\ldots,p$. The symbols $\theta(j_\sigma,k_\sigma)$ keep track of labels which swap their relative positions in the two permutations $\sigma$ and $\rho$, i.e. $\theta(j_\sigma,k_\sigma)=1$ (=0) if the ordering of the legs $j_\sigma,k_\sigma$ is the same (opposite) in the ordered sets $\sigma(2,3,\ldots,p)$ and $\rho(2,3,\ldots,p)$. The inverse in (\ref{BGdef}) is taken w.r.t.~matrix multiplication which treats $\sigma$ and $\rho$ as row- and column indices.} $S[\cdot | \cdot ]_1$ \cite{momKern},
\beq
\label{BGdef}
{\cal K}_{1\sigma(23\ldots p)} \equiv \sum_{\rho \in S_{p-1}} S^{-1}[\sigma|\rho]_1 \, K_{1\rho(23\ldots p)} 
\ ,
\eeq
with permutation $\sigma \in S_{p-1}$ of legs $2,3,\ldots,p$.

The combination of color-ordered trees as in (\ref{BGexpl}) and (\ref{BGdef}) simplifies
their multiparticle equations of motion \cite{EOMBBs}
\begin{align}
\label{nabla8}
\big\{ D_{(\alpha} , {\cal A}_{\beta)}^B \big\} &=
\g^m_{\alpha \beta} {\cal A}_m^B + \! \! \sum_{XY=B} \! \! \bigl( {\cal A}_\alpha^X {\cal A}_{\beta}^Y
-{\cal A}_\alpha^Y {\cal A}_{\beta}^X\bigr)
\\
\bigl[ D_\alpha ,{\cal A}_m^B\bigr] &= k^B_m {\cal A}_{\alpha}^B + (\g_m {\cal W}_B)_\alpha + \!\! \!
\sum_{XY=B} \! \! \bigl( {\cal A}_\alpha^X {\cal A}_{m}^Y -{\cal A}_\alpha^Y {\cal A}_{m}^X\bigr)
\cr
\big\{ D_{\alpha} ,{\cal W}^\beta_B \big\} &= \tfrac{1}{4} (\g^{mn})_{\alpha}{}^{\beta} {\cal F}^B_{mn} 
+ \! \! \sum_{XY=B}  \! \!\bigl( {\cal A}_\alpha^X {\cal W}_Y^{\beta} -{\cal A}_\alpha^Y {\cal W}_X^\beta\bigr)
\cr
\big[D_{\alpha}, {\cal F}^{mn}_B \big] &= k_B^{[m} (\g^{n]} {\cal W}_B)_\alpha 
+\!\! \sum_{XY=B} \!\!\bigl( {\cal A}_\a^X {\cal F}_Y^{mn} -{\cal A}_\alpha^Y {\cal F}_X^{mn}\bigr) \notag \\
  & +\! \!\sum_{XY=B}\! \! \bigl( {\cal A}_X^{[n}  (\g^{m]} {\cal W}_Y)_\alpha - {\cal A}_Y^{[n}
  (\g^{m]} {\cal W}_X)_\a \bigr) \ .
\notag
\end{align}
Momenta $k_{B} \equiv k_1+k_2+\ldots k_p$ are associated with multiparticle labels $B=12\ldots p$,
and $\sum_{XY=B}$ instructs to sum over all their deconcatenations into non-empty $X=12\ldots j$ and $Y=j+1\ldots p$
with $1\leq j \leq p-1$. E.g.~the three-particle equation of motion
of $\cA_\a^{123}$ reads
\begin{align}
\label{simpler}
\big\{ D_{(\alpha} , {\cal A}_{\beta)}^{123} \big\} &=
\g^m_{\alpha \beta} {\cal A}_m^{123}\\
&+ {\cal A}_\alpha^1 {\cal A}_{\beta}^{23}
+ {\cal A}_\alpha^{12} {\cal A}_{\beta}^3
-{\cal A}_\alpha^{23} {\cal A}_{\beta}^1
-{\cal A}_\alpha^3 {\cal A}_{\beta}^{12} \ ,\notag
\end{align}
and a comparison with \eqref{DAthree} highlights the advantages of the diagram expansions in
\eqref{BGexpl}.
Superfields up to multiplicity five satisfying \eqref{nabla8}
were explicitly constructed in \cite{EOMBBs} and there
are no indications of a breakdown of \eqref{nabla8}
at higher multiplicity.

The symmetry properties of the ${\cal K}_B$ can be inferred from their cubic-graph expansion and
summarized as \cite{Berends:1988zn}
\beq
\label{alternal}
{\cal K}_{A\shuffle B} = 0,\quad \forall A,B\neq\emptyset \ ,
\eeq
where $\shuffle$ denotes the shuffle product\footnote{The shuffle product in ${\cal K}_{A\shuffle B}$ is
defined to sum all ${\cal K}_{\sigma}$ for permutations $\sigma$ of $A \cup B$ which preserve the
order of the individual elements of both sets $A$ and $B$.} \cite{Ree}. For example,
\begin{align}
0&= {\cal K}_{1 \shuffle 2} = \cK_{12} + \cK_{21} \notag \\
0&={\cal K}_{1 \shuffle 23} =
 \cK_{123} + \cK_{213} + \cK_{231}
\\
0&={\cal K}_{12 \shuffle 3} -{\cal K}_{1 \shuffle 32}  =
\cK_{123}  -  \cK_{321}  \,,
\notag
\end{align}
and symmetries (\ref{alternal}) at higher multiplicity leave $(p-1)!$ independent
permutations of ${\cal K}_{12\ldots p}$. Any permutation can be expanded in a
basis of ${\cal K}_{1\sigma(23\ldots p)}$ with $\sigma \in S_{p-1}$ through
the Berends--Giele symmetry
\beq
{\cal K}_{B1A} = (-1)^{|B|} {\cal K}_{1(A\shuffle B^t)} \ ,
\label{KlKu}
\eeq
where $|B|=p$ and $B^t = b_p \ldots b_2 b_1$ for a multi particle label $B=b_1 b_2\ldots b_p$.
Since the Berends--Giele current ${\cal K}_{12…p}$ is composed from the cubic diagrams in a partial
amplitude with an additional off-shell leg, \eqref{KlKu} can be understood as a Kleiss--Kuijf relation among
the latter \cite{Kleiss:1988ne}.

\section{\label{sec:lie} Generating series of SYM superfields}

In order to connect multiparticle fields and Berends--Giele currents with
the non-linear field equations (\ref{EOMs}), we define generating
series ${\mathbb K} \in \{  \bA_\alpha, \bA^m , \bW ^\alpha, \bF^{mn} \}$
\begin{align}
\label{Lie}
{\mathbb K} &\equiv \sum_i {\cal K}_i t^i + \sum_{i,j} {\cal K}_{ij} t^i t^j + \sum_{i,j,k} {\cal
K}_{ijk} t^i t^j t^k+ \ldots\\
&= \sum_i \cK_i t^i + \tfrac{1}{2}\sum_{i,j}\cK_{ij}[t^i,t^j] +\tfrac{1}{3}\sum_{i,j,k}\cK_{ijk}[[t^i,t^j],t^k] + \cdots
\notag
\end{align}
where $t^i$ denote generators in the Lie algebra of the non-abelian gauge group. Hence, the generating series in (\ref{Lie}) adjoin color degrees of freedom to the polarization and momentum dependence in the linearized multiparticle superfields ${\cal K}_B$. The second line follows from the symmetry \eqref{alternal}, which guarantees that
$\mathbb K$ is a Lie element \cite{Ree}.

As a key virtue of the series ${\mathbb K} \in \{  \bA_\alpha, \bA^m , \bW ^\alpha, \bF^{mn} \}$ in (\ref{Lie}), they allow to rewrite the $D_\alpha$ action on Berends--Giele currents ${\cal K}_B \in \{  {\cal A}^B_\alpha, {\cal A}_B^m , {\cal W}_B^\alpha, {\cal F}_B^{mn} \}$ in (\ref{nabla8})
as non-linear equations of motion,
\begin{align}
 \big\{ D_{(\a} , \bA_{\beta)} \big\} &= \g^m_{\a \beta} \bA_m + \{ \bA _\a ,\bA_\beta\}\cr
\big[ D_\a , \bA_m\big] &= \big[ \partial_m ,\bA_{\a} \big] + (\g_m \bW)_\a + [\bA_\a,\bA_m]\cr
\big\{ D_{\a} ,\bW^\beta \big\} &= \tfrac{1}{4} (\g^{mn})_{\a}{}^{\beta} \bF_{mn} + \{\bA_\a, \bW^\b\}\cr
\big[D_{\a}, \bF^{mn} \big] &= \big[ \partial^{[m} ,(\g^{n]} \bW)_\a \big] + [ \bA_\a, \bF^{mn}] \notag\\
& -  [\bA^{[m}, (\g^{n]} \bW)_\a]\, \ ,
\label{bbEOMs}
\end{align}
where $[\partial^{m},{\mathbb K}]$ translates into components $k_B^m{\cal K}_B$.

Remarkably, they are equivalent to the non-linear SYM field equations \eqref{EOMs} if the connection in
(\ref{covder}) is defined through the representatives $\bA_\alpha$ and $\bA_m$ of the generating series
in (\ref{Lie}). In other words, the resummation of linearized multiparticle superfields $\{ A^B_\alpha,
A_B^m , W_B^\alpha, F_B^{mn} \}$ through the generating series (\ref{Lie}) of Berends--Giele currents
(\ref{BGdef}) solves the non-linear SYM equations (\ref{EOMs}).

Given that the multiparticle superfields satisfy \cite{EOMBBs}
\begin{gather}
{\cal F}_B^{mn} = k^{[m}_B {\cal A}^{n]}_B - \! \! \sum_{XY=B} \! \! ( {\cal A}_X^{m} {\cal A}^{n}_Y-{\cal A}_Y^{m} {\cal A}^{n}_X)
 \label{nabla23} \\
k_m^B (\g^m {\cal W}_B)_\alpha =\! \!  \sum_{XY=B} \! \! \big[ {\cal A}_m^X (\g^m
{\cal W}_Y)_\alpha- {\cal
A}_m^Y (\g^m {\cal W}_X)_\alpha\big]\cr
k_m^B {\cal F}^{mn}_B = \! \!  \sum_{XY=B} \! \! \bigl[  2({\cal W}_X \g^n {\cal W}_Y) 
+ {\cal A}_m^X {\cal F}^{mn}_Y- {\cal A}_m^Y {\cal F}^{mn}_X\bigr]\,,
\notag
\end{gather}
the above definitions are compatible with (\ref{FS}) and
\beq
\label{extra}
\big[ \nabla_m , (\g^m \bW)_\alpha \big] = 0\,,\quad
\big[ \nabla_m, \bF^{mn} \big]=  \g^n_{\alpha \beta} \big\{ \bW^\alpha, \bW^\beta\big\} \,,
\eeq
whose lowest components in $\theta$ encode the Dirac and Yang-Mills equations for the gluino and gluon field.

A linearized gauge transformation in particle one,
\begin{align}
\delta_1 A_\alpha^1 = D_{\alpha}\Omega_1 ,\quad \delta_1 A_m^1 = k_m^1 \Omega_1 ,\quad 
\label{lingauge}
\end{align}
can be described by a scalar superfield $\Omega_1$, it shifts the gluon polarization by its momentum. The gluino component and the linearized field strengths are invariant under (\ref{lingauge}), $\delta_1 W_1^\alpha = \delta_1 F_1^{mn}=0$, whereas Berends--Giele currents of multiparticle superfields with $B=12\ldots p$ transform as follows \cite{Mafra:2014gsa}:
\begin{align}
\delta_1 {\cal A}^B_\alpha &= \big[D_\alpha ,\Omega_B \big] +\! \! \! \sum_{XY=B}\! \!\Omega_X{\cal A}^Y_\alpha, \quad
\delta_1 {\cal W}_B^\alpha = \! \! \! \sum_{XY=B} \! \! \Omega_X{\cal W}_Y^\alpha
\notag \\
\delta_1 {\cal A}^B_m &= \big[\partial_m,\Omega_B \big] + \! \! \!\sum_{XY=B} \! \!\Omega_X{\cal A}^Y_m, \quad
\delta_1 {\cal F}_B^{mn} = \! \! \! \sum_{XY=B} \! \!\Omega_X{\cal F}_Y^{mn}   .
\label{multgauge}
\end{align}
The multiparticle gauge scalars $\Omega_{12\ldots p}$ are exemplified in
appendix B of \cite{Mafra:2014gsa} and gathered in the generating series
\beq
\mathbb L_1 \equiv   \Omega_1 t^1 + \sum_i  \Omega_{1i}[t^1,t^i] + \sum_{j,k} \Omega_{1jk} [ [t^1, t^j],t^k ] + \ldots 
\eeq
This allows to cast (\ref{multgauge}) into the standard form of non-linear gauge transformations,
\begin{align}
\delta_1 \bA_\alpha &
   = \big[\nabla_\alpha ,\mathbb L_1 \big] , \quad \ \, \delta_1 \bW^\alpha =  \big[\mathbb L_1,\bW^\alpha  \big] 
\label{gau5}
\\
\delta_1 \bA_m &
 = \big[\nabla_m ,\mathbb L_1 \big] ,\quad \delta_1 \bF^{mn} =  \big[\mathbb L_1,\bF^{mn}  \big]   \ ,
\notag
\end{align}
such that traces w.r.t.~generators $t^i$ furnish a suitable starting point to construct gauge invariants.

\section{\label{sec:higherdim} Higher mass dimension superfields}

The introduction of the Lie elements ${\mathbb K}$ and their associated
supercovariant derivatives allow the recursive definition of
superfields with higher mass dimensions,
\begin{align}
\bW^{m_1\ldots m_k\alpha} &\equiv \big[ \nabla^{m_1} , \bW^{m_2\ldots
m_k\alpha}\big]\,,\label{highmass}\\
\bF^{m_1\ldots m_k|pq} &\equiv \big[ \nabla^{m_1} , \bF^{m_2\ldots m_k|pq} \big]\,,\notag
\end{align}
where the vertical bar separates the antisymmetric pair of indices present in the
recursion start $\bF^{pq}$. Their component fields are defined by
\begin{align}
\bW^{m_1\ldots m_k\alpha} &\equiv
 \sum_{B\neq \emptyset} t^B {\cal \cW}_{B}^{m_1\ldots m_k\alpha}\,,\label{defbWbF}\\
\bF^{m_1\ldots m_k|pq} &\equiv  \sum_{B\neq \emptyset} t^B \cF_{B}^{m_1\ldots m_k|pq}\,,\notag
\end{align}
with $t^B \equiv t^1 t^2 \ldots t^p$ for $B=12\ldots p$. They inherit the Berends--Giele
symmetries (\ref{alternal}) and can be identified as
\begin{align}
 {\cal W}_{B}^{m_1\ldots m_k\alpha} &= k_B^{m_1} {\cal W}_{B}^{m_2\ldots m_k\alpha}\label{bcomp}\\
&+\!\! \sum_{XY=B}\!\! \bigl(  {\cal W}_{X}^{m_2\ldots m_k \alpha} {\cal A}^{m_1}_Y
-{\cal W}_{Y}^{m_2\ldots m_k \alpha} {\cal A}^{m_1}_X\bigr)\,,\cr
{\cal F}_{B}^{m_1\ldots m_k|pq} &= k_B^{m_1} {\cal F}_{B}^{m_2\ldots m_k|pq} \cr
&+\!\! \sum_{XY=B}\!\! \bigl(  {\cal F}_{X}^{m_2\ldots m_k|pq} {\cal A}^{m_1}_Y-{\cal F}_{Y}^{m_2\ldots m_k|pq}
{\cal A}^{m_1}_X\bigr)\,.\notag
\end{align}
Note from \eqref{bcomp} that the non-linearities in the definition of higher mass superfields
do not contribute in the single-particle context with
${\cal W}_{i}^{m_1\ldots m_k\alpha} = k_i^{m_1} {\cal W}_{i}^{m_2\ldots m_k\alpha}$ whereas
the simplest two-particle correction reads
\beq
{\cal W}_{12}^{m\alpha} = k_{12}^m {\cal W}^{\alpha}_{12} + {\cal W}_1^\alpha {\cal A}_2^m- {\cal W}_2^\alpha {\cal A}_1^m
\ .
\eeq


\medskip
\noindent
{\bf Equations of motion at higher mass dimension:} 
The equations of motion for the superfields of higher mass dimension \eqref{highmass}
follow from $\big[ \nabla_\alpha,\nabla_m \big] = -(\g_m \bW)_\alpha$ and
$\big[ \nabla_m,\nabla_n \big] = - \bF_{mn}$ together with
Jacobi identities among iterated brackets.
The simplest examples are given by
\begin{align}
\big\{ \nabla_\alpha ,\bW^{m\beta} \big\} &=
\tfrac{1}{4} (\g_{pq})_\alpha{}^\beta \bF^{m|pq} - \big\{ ( \bW\g^m)_\alpha, \bW^\beta\big\}\,,\cr
\big[ \nabla_\alpha , \Fg^{m|pq} \big] &=( \Wg^{m[p} \g^{q]})_\alpha
- \big[ (\Wg \g^m)_\alpha, \Fg^{pq} \big]\,,
\end{align}
which translate into component equations of motion
\begin{align}
D_\alpha {\cal W}^{m\beta}_{B} &= \tfrac{1}{4} (\g_{pq})_\alpha{}^\beta {\cal F}_{B}^{m|pq}
+\!\! \sum_{XY=B} \!\!({\cal A}^X_{\alpha} {\cal W}^{m\beta}_{Y} -{\cal A}^Y_{\alpha} {\cal W}^{m\beta}_{X}) \notag \\
& -\!\!\sum_{XY=B}\!\! \big[ ( {\cal W}_X \g^m)_\alpha {\cal W}^{\beta}_{Y} -
( {\cal W}_Y \g^m)_\alpha {\cal W}^{\beta}_{X}
\big]\,,
\notag \\
D_\alpha {\cal F}^{m|pq}_{B} &=  ({\cal W}_{B}^{m[p} \g^{q]})_{\a} 
+\!\!\sum_{XY=B}\!\! ({\cal A}^X_{\alpha} {\cal F}^{m|pq}_{Y} 
- {\cA}^Y_{\a} \cF^{m|pq}_{X}) \notag \\
& -\!\! \sum_{XY=B}\!\! \big[ ( {\cal W}_X \g^m)_\a \cF^{pq}_{Y} -
( {\cal W}_Y \g^m)_\alpha {\cal F}^{pq}_{X}
\big]\, \!.
\label{nabla41b}
\end{align}
In general, one can prove by induction that
\begin{align}
\big\{ \nabla_\alpha , \Wg^{N\beta} \big\} &= \tfrac{1}{4} (\g_{pq})_\alpha{}^{\beta} \Fg^{N|pq}
-\!\!\!\! \sum_{M \in P(N) \atop{M \neq \emptyset}}\!\!\!\!\! \big\{ (\Wg \g)^M_\alpha , \Wg^{(N\setminus M) \beta} \big\}
\cr
\big[ \nabla_\alpha , \Fg^{N|pq} \big] &=  ( \Wg^{N[p} \g^{q]})_\alpha
- \!\!\!\sum_{M \in P(N) \atop{M \neq \emptyset}}\!\!\! \big[ (\Wg \g)^M_\alpha , \Fg^{(N\setminus M)
pq} \big]  \, \! .\label{nabla47}
\end{align}
The vector indices have been gathered to a multi-index $N\equiv n_1n_2\ldots n_{k}$. Its
power set $P(N)$ consists of the $2^{k}$ ordered subsets, and
$(\Wg \gamma)^N\equiv (\Wg^{n_1\ldots n_{k-1}} \gamma^{n_{k}})$.

The higher-mass-dimension superfields obey further relations which
can be derived from Jacobi identities of nested (anti)commutators. For example,
(\ref{FS}) determines
their antisymmetrized components
\begin{align}
\Wg^{[n_1 n_2] n_3\ldots n_k \beta}_{}&= \big[  \Wg^{n_3\ldots n_k \beta} , \Fg^{n_1n_2} \big]
 \label{nabla50} \\
\Fg^{[n_{1} n_{2}] n_{3} \ldots n_k |pq} &= \big[  \Fg^{n_3\ldots n_k |pq} , \Fg^{n_1n_2} \big]\,.
\notag
\end{align}
Moreover, the definitions (\ref{highmass}) via iterated commutators imply that
\begin{align}
\label{lie3}
\Fg^{[m|np]} &= 0, \quad \Fg^{[mn]|pq} + \Fg^{[pq]|mn}  = 0 \,,
\end{align}
and the gauge-variations $\delta_1 \nabla_m =  [\mathbb L_1,\nabla_m]$ and (\ref{gau5}) yield
\beq
\delta_1 \bW^\alpha_{N} =  \big[\mathbb L_1,\bW^\alpha_{N}  \big] 
, \quad 
\delta_1 \bF^{N|pq} =  \big[\mathbb L_1,\bF^{N|pq}  \big]  \ .
\eeq
Manifold generalizations of (\ref{extra}), (\ref{nabla50}) and (\ref{lie3}) can be generated using these
same manipulations.

\section{\label{sec:outline} Outlook and applications}

The representation of the non-linear superfields of ten-dimensional SYM theory described in
this letter was motivated by the computation of scattering amplitudes in the pure spinor formalism.
Accordingly, they give rise to generating functions for amplitudes. For
example, color-dressed SYM amplitudes at tree-level $M(1,2,\ldots,n)$ involving particles
$1,2,\ldots,n$ are generated by
\beq
\label{treeEx}
\frac{1}{3}\tr \langle \PsiF \PsiF \PsiF \rangle =
\sum_{n=3}^{\infty}(n-2) \! \! \! \! \! \! \sum_{i_{1}<i_2<\ldots <i_n}\!\!\!\! \! \! M(i_1,i_2,\ldots,i_n)  \ .
\eeq
As firstly pointed out in the appendix of \cite{Grassi:2011ky}, the generating
series $\PsiF \equiv \lambda^\alpha \bA_\alpha$ involving the pure spinor $\l^\a$
satisfies the field equations $Q\PsiF= \PsiF \PsiF$ of the action $\tr\int {\rm d}^{10}x \langle \tfrac{1}{2}\PsiF Q \PsiF 
- \tfrac{1}{3}\PsiF\PsiF\PsiF\rangle$ \cite{sparticle} with BRST operator $Q\equiv \lambda^\alpha D_\alpha$.
The zero mode prescription of
schematic form $\langle \lambda^3 \theta^5\rangle=1$ extracting the gluon- and gluino components
is explained in \cite{psf} and automated in \cite{Mafra:2010pn}. With $n=3$ or $n=4$ external
states, for instance, assembling the components $V^B \equiv \lambda^\alpha A^B_\alpha$ of
$\PsiF \PsiF \PsiF$ in (\ref{treeEx}) with the appropriate number of labels yields
\begin{align}
M(1,2,3) &= \tr(t^1 [t^2,t^3])  \langle V^1 V^2 V^3 \rangle \\
2M(1,2,3,4) &= \tr(t^1 t^2 t^3t^4)  \Big \langle \frac{V^{12} V^3 V^4}{s_{12}} + \frac{V^{23} V^4 V^1}{s_{23}} \notag \\
& \!\!  + \frac{V^{34} V^1 V^2}{s_{34}} + \frac{V^{41} V^2 V^3}{s_{41}} \Big \rangle  + {\rm perm}(2,3,4) \ .\notag
\end{align}
The pure spinor representation of ten-dimensional\footnote{
See \cite{SYMfourD} for expressions in $D=4$ upon
specification of helicities.}
$n$-point SYM amplitudes is described in
\cite{npttree}. Further
details and generalizations to superamplitudes with a single insertion of supersymmetrized operators
$F^4$ or $D^2F^4$ will be given elsewhere \cite{wip}.

The multiparticle superfields of higher mass dimensions can be used to obtain
simpler expressions for higher-loop kinematic factors of superstring amplitudes. For example,
the complicated three-loop kinematic factors
generating the matrix element of the (supersymmetrized) operator $D^6R^4$ \cite{3loop}
\beq
\label{Dsix}
M_{D^6 R^4} = {| T_{12,3,4} |^2\over s_{12}} + |T^m_{1234}|^2
+ (1,2|1,2,3,4)
\eeq
can be equivalently represented by
\begin{align}
T_{12,3,4} &\equiv \langle (\l\g_m W_{12}^n)(\l\g_n W_{[3}^p)(\l\g_p W_{4]}^m)\rangle
\notag \\
T^m_{1234} &\equiv \langle A^m_{(1} T_{2),3,4}^{\phantom m}
+(\l\g^m W_{(1}^{\phantom m}) L^{\phantom m}_{2),3,4}\rangle \label{new3loop} \\
L_{2,3,4} &\equiv \tfrac{1}{3}(\l\g^n W_{[2}^q)(\l\g^q W_{3}^p)F_{4]}^{np}\rangle \ .
\notag
\end{align}
In \eqref{Dsix}, the notation $(A_1,\ldots , A_p \mid A_1,{\ldots} ,A_n)$ instructs to sum over all possible ways to choose $p$ elements $A_1,A_2,\ldots,A_p$ out of the set $\{A_1,{\ldots} ,A_n\}$, for a total of
${n\choose p}$ terms.
The tensor products of left- and right-moving SYM superfields in $|T_{12,3,4}|^2=T_{12,3,4}\tilde T_{12,3,4}$ are understood to yield superfields of type IIB or type IIA supergravity. Accordingly, the component polarizations of the supergravity multiplet arise from the tensor product of gluon polarizations and gluino wavefunctions within the SYM superfields.

The low-energy limit of the three-loop closed string amplitude given in (\ref{Dsix}) is proportional to
the $(\alpha')^6$ correction of the corresponding tree-level amplitude which in turn defines the
aforementioned $D^6R^4$ operator.

It would be interesting to explore the dimensional reduction \cite{SYM} of the above setup and its generalization to SYM theories with less supersymmetry or to construct formal solutions to supergravity
field equations along similar lines. 

\medskip
\noindent\textit{Acknowledgements:} We thank Eduardo Casali, Evgeny Skvortsov, Massimo Taronna and
Stefan Theisen for valuable discussions as well as Nathan Berkovits for helpful comments on the draft. 
This research
has received funding from
the European Research Council under the European Community's Seventh Framework Programme
(FP7/2007-2013)/ERC grant agreement no. [247252].

\end{document}